\documentclass[twocolumn,showpacs,preprintnumbers,amsmath,amssymb,fleqn]{revtex4} 
\usepackage{amsmath,amssymb,graphicx}
\begin{document}

\title{Enhancing quadratic optomechanical coupling via a nonlinear medium and lasers}

\author{Jian-Song Zhang$^{1}$}
\author{Ming-Cui Li$^{1}$}
\author{Ai-Xi Chen$^{2,3}$}
\email{aixichen@zstu.edu.cn}
\affiliation{$^{1}$Department of Applied Physics, East China Jiaotong University,
Nanchang 330013, People's Republic of China \\
$^{2}$ Department of Physics, Zhejiang Sci-Tech University, Hangzhou 310018, People's Republic of China\\
$^{3}$ Institute for Quantum Computing, University of Waterloo, Ontario N2L3G1, Canada}

\begin{abstract}
We propose a scheme to significantly increase quadratic optomechanical couplings of optomechanical systems with the help of a nonlinear medium and two driving lasers. The nonlinear medium is driven by one laser and the optical cavity mode is driven by a strong laser. We derive
an effective Hamiltonian using squeezing transformation and rotating wave approximation.
The effective quadratic optomechanical coupling strength can be larger than the decay rate of the cavity mode by adjusting the two optical driving fields. The thermal noise of squeezed cavity mode can be suppressed totally with the help of a squeezed vacuum field.
Then, a driving field is applied to the mechanical mode. We investigate the equal-time second-order correlations and find there are photon,  phonon, and photon-phonon blockades even when the original single-photon quadratic coupling is much smaller than the decay rate of the optical mode. In addition, the sub-Poissonian window of the two-time second-order correlations can be controlled by the mechanical driving field. Finally, we show the
squeezing and entanglement of the model could be tuned by the driving fields of the nonlinear medium and mechanical mode.

\end{abstract}
\maketitle

\section{Introduction}
In recent years, considerable efforts have been devoted to the study of optomechanical systems
since they have a wide range of applications including highly sensitive measurement of tiny displacement,
creation of nonclassical states of light and mechanical motion, and quantum information processing \cite{Aspelmeyer2014,Bowen2015,Xiong2015,Yin2017,Xiong2017}.
Single-photon emitters play an important role in quantum information processing \cite{Aspelmeyer2014,Bowen2015,Xiong2015}.
In experiments, single-photon emitters are usually implemented by the so-called ``photon blockade" effect \cite{Aspelmeyer2014}.
The interactions between the cavity mode and mechanical mode changes the energy spectrum of the system and
an anharmonic ladder in energy spectrum is formed. Thus, the first photon in an optomechanical system blocks the
transmission of a second photon.

A typical quantum cavity optomechanical system is consisted of a movable mirror and an optical cavity where the radiation pressure
is nonlinear. In order to enter the quantum nonlinear region, the single-photon coupling strength $g_0$ should be comparable to
the decay rate of the cavity mode $\kappa$ and the frequency of the mechanical mode $\omega_m$ \cite{Lemonde2016}.
Unfortunately, the single-photon coupling constant $g_0$ is, in general, much smaller than the decay rate of cavity $\kappa$
and the mechanical mode frequency $\omega_m$ \cite{Aspelmeyer2014,Bowen2015,Xiong2015}. Therefore, it is highly desirable to
propose schemes to enhance the coupling strength in current experiments.
One way to enhance the optomechanical coupling constant is to apply a strong coherent field on the cavity mode \cite{Lemonde2013,Lemonde2015,Komar2013,Ludwig2012,Liao2015,Xuxunnong2015,Chan2012}. However, the coupling constant
is still much smaller than the frequency of the mechanical mode.

Recently, the ``unconventional photon blockade" (UPB) scheme was proposed by Liew and Savona \cite{Liew2010}.
In the UPB scheme, there is photon blockade effect even when the nonlinear energy shifts and driving field are both small
due to the quantum interferences between different excitation pathways. However, there is one obstacle in the original UPB scheme.
The large coupling induced by a weak nonlinearity makes the second-order correlations oscillate rapidly \cite{Liew2010,Bamba2011,Flayac2013,Flayac2016,Xu2014,Flayac2017}.
The single-photon regime is defined by $g^{(2)}(\tau) < 0.5$ with $g^{(2)}(\tau)$ being the delayed
two-photon correlation function \cite{Flayac2016}. In the original UPB, in order to enter the single-photon regime,
the delayed time $\tau$ should be smaller than the cavity lifetime $1/\kappa$, i.e., $\tau < 1/\kappa$.
Thus, it is difficult to enter the single-photon regime in experiments since the sub-Poissionian window of the UPB scheme is very small.
This obstacle can be overcome by employing a mutual driving of the modes and a mixing of the output \cite{Flayac2017}.
Here, we overcome this obstacle by using a mechanical driving field as we shall see in next section.

The influence of nonlinear media on the dynamics of optomechanical systems was extensively investigated\cite{Law1995,Xuererb2012,Huang2009,Agarwal2013,Zhang2017,Lv2015}.
The movable mirror of an optomechanical system can be efficiently cooled to about millikelvin temperature
with the help of the nonlinear medium \cite{Huang2009}. The normal mode splitting, cooling,
and entanglement in optomechanical systems can be tuned by the nonlinear medium \cite{Zhang2017}.
In Ref. \cite{Lv2015}, the authors have studied the nonlinear interaction of a mechanical mode and a squeezed cavity mode in
optomechanics with nonlinear media. They found that the effective coupling strength of optomechanical systems can be about three orders of
the magnitude larger than the original single-photon coupling strength. Thus, the nonlinear interaction between the
optical cavity mode and mechanical mode can be significantly enhanced by employing the nonlinear media and driving field \cite{Lv2015}.

Generally, one can classify the optomechanical couplings into two classes. One is the linear coupling class where the coupling
strength is proportional to $q$ with $q$ being the displacement of the mechanical oscillator.
The other is the quadratic coupling class where the coupling strength is proportional to $q^2$.
We would like to point out that the scheme proposed in Ref. \cite{Lv2015} is used to enhance the coupling strength
of linear optomechanics. In particular, a mechanical driving field is introduced to cancel the force induced by
the parametric amplification in deriving the effective Hamiltonian of Eq.(2) in Ref. \cite{Lv2015}.
Consequently, the mechanical driving field disappears in the effective Hamiltonian.

Many efforts have been invested in the single-photon strong-coupling regime of linear optomechanics
both theoretically \cite{Rabl2011,Nunnenkamp2011,Hong2013,Liao2012,He2012,Xu2013,Kronwald2013,Liao20131,Lv2013,Hu2015,Xie2016} and experimentally \cite{Gupta2007,Murch2008,Brennecke2008,Eichenfield2009}.
In recent years, much attention has been paid to the study of quadratic optomechanical systems \cite{Thompson2008,Bhattacharya2008,Bhattacharya20082,Rai2008,Jayich2008,Sankey2010,Nunnenkamp2010,Liao2013,Liao2014,Xie2017,Xu2018,Zhu2018,Lee2018}.
A few methods have been suggested to enhance the quadratic coupling strength of optomechanical systems \cite{Vanner2011,Jacobs2012,Yin20182}.
In Ref. \cite{Vanner2011}, a measurement-based method was proposed to select linear or quadratic optomechanical
coupling and obtain an effective quadratic optomechanics.
It has been demonstrated experimentally that the quadratic coupling constant can be remarkably increased by using a fiber cavity \cite{Jacobs2012}. Very recently, the authors of Ref. \cite{Yin20182} proposed a scheme to enhance photon-phonon cross-Kerr nonlinearity
via two-photon driving fields.

In the present paper, we propose a scheme to significantly enhance the quadratic coupling strength via
nonlinear media and two lasers. The nonlinear media are put into the cavity optomechanics system and
pumped by a laser which is an optical parametric amplifier (OPA). The optical cavity is driven by another strong driving field.
Using the squeezing transformation, the Hamiltonian of the system can be transferred to a standard
optomechanical system with quadratic coupling. Different from the linear coupling case, the parametric amplification
changes the frequency of the mechanical mode in the quadratic optomechanics.
Therefore, it is not necessary to introduce a mechanical driving field
in the derivation of the Hamiltonian corresponding to the standard quadratic
optomechanical system (see the next section for more details).
The total amplification of the quadratic coupling strength depends on two factors.
One is the strong optical driving field used to
improve the quadratic coupling strength between the optical mode and mechanical mode effectively.
The other is the OPA which can also increase the quadratic coupling constant.
Furthermore, the total amplification is the product of the two factors mentioned above.
Thus, the effective coupling strength can be significantly increased with the help of nonlinear medium and two driving lasers.

Then, we study the photon blockade, phonon blockade, photon-phonon blockade, squeezing, and entanglement in the present model by applying a mechanical driving field. We find the photon blockade and phonon blockade can be observed in the same parameter regime.
The sub-Poissonian window of the two-time second-order correlations can be controlled by the mechanical driving field
and the delayed time $\tau$ could be much larger than the lifetime of the cavity $1/\kappa$.
This is very important in experiments. The squeezing of the optical and mechanical modes can also be tuned by
the driving fields. In order to investigate the entanglement of the system,
we adopt criteria introduced by Duan \emph{et al.} and Hillery and Zubairy.
Our results show that Duan's criterion is not able to detect the entanglement of the optical and mechanical modes
while the Hillery-Zubairy criterion is able to detect the entanglement in the present model.
There is stationary entanglement between optical and mechanical modes which can be controlled by the driving fields.

The organization of this paper is as follows. In Sec. II, we introduce the model and
derive an effective Hamiltonian using the squeezing transformation and rotating wave approximation.
In Sec. III, we discuss the effects of the nonlinear media and driving fields on the second-order correlations of the present model.
In Sec. IV, we investigate the influence of nonlinear media on the squeezing of the quadratic optomechanics.
In Sec. V, the entanglement of the model is investigated by using the criteria introduced by \emph{et al.} and Hillery and Zubairy.
In Sec. VI, we summarize our results.

\section{Model and Hamiltonian}
\subsection{Hamiltonian}
A Fabry-Perot cavity is formed by two mirrors within a thermal bath. One mirror is partially transparent and
the other mirror is perfectly reflecting. A thin partially reflecting membrane is put into the optical cavity.
Then, a nonlinear medium is also put into the cavity and is pumped by a field with driving frequency $2\omega_L$, amplitude
$\lambda/2$, and phase $\phi_L$. In the present work, we consider a quadratic optomechanical system where an
optical mode is quadratically coupled to a mechanical mode. This kind of coupling can be found in
cavities with a membrane-in-the-middle system \cite{Huang2009,Liao2013,Liao2014,Xie2017,Xu2018} and other optomechanical systems \cite{Aspelmeyer2014,Bowen2015}.
The Hamiltonian of the system is ($\hbar = 1$) \cite{Huang2009,Xu2018,Xie2017,Liao2013,Liao2014}
\begin{eqnarray}
H_0 &=& H_C + H_m + H_I,\\
H_c &=& \omega_c A^{\dag} A + \frac{\lambda}{2}(e^{-2i\omega_L t - i\phi_L} A^{\dag 2} + e^{2i\omega_L t + i\phi_L} A^{ 2}),\\
H_m &=& \frac{\omega_m}{2}(Q^2 + P^2) = \omega_m B^{\dag}B,\\
H_I &=& 2g_0A^{\dag}A Q^2 = g_0 A^{\dag}A(B^{\dag} + B)^2,
\end{eqnarray}
where $A$ and $A^{\dagger}$ are the annihilation and creation operators of the cavity mode with frequency $\omega_{c}$, and
$Q$ and $P$ are, respectively, the dimensionless position and momentum operators of the movable mirror with frequency $\omega_m$.
Here, the dimensionless position $Q$ and momentum $P$ are defined by $Q = (B^{\dag} + B)/\sqrt{2}$
and $P = i(B^{\dag} - B)/\sqrt{2}$
where $B$ and $B^{\dagger}$ are, respectively, the annihilation and creation operators of the mechanical mode with frequency $\omega_m$.
Here, $q$ and $p$ are the position and momentum of the membrane, respectively.
$g_0$ is the single-photon quadratic coupling strength and $\lambda/2$ is the amplitude of the driving field.
The second term in $H_c$ corresponds to the OPA of the system.

In the rotating reference frame defined by $U_1 = \exp{(-i\omega_L t A^{\dag}A)}$ the Hamiltonian $H_0$ can be rewritten as
\begin{eqnarray}
H_0 &=& \Delta_c A^{\dag}A + \frac{\lambda}{2}(e^{-i\phi_L} A^{\dag 2} + e^{i\phi_L} A^2) + \frac{\omega_m}{2} (Q^2 + P^2)\nonumber\\
&& + 2g_0 A^{\dag}A Q^2,
\end{eqnarray}
where $\Delta_c = \omega_c - \omega_L$.
We first diagonalize the Hamiltonian $H_C = \Delta_c A^{\dag}A + \frac{\lambda}{2}(e^{-i\phi_L} A^{\dag 2} + e^{i\phi_L} A^2)$ using the squeezing transformation \cite{Lemonde2015,Lv2015} defined by $A = \cosh(r) \widetilde{A} - e^{-i\phi_L} \sinh(r) \widetilde{A}^{\dag}$.
Then, the Hamiltonian $H_c$ is transformed into
\begin{eqnarray}
\widetilde{H}_c = \widetilde{\Delta}_c \widetilde{A}^{\dag}\widetilde{A},
\end{eqnarray}
with $\widetilde{\Delta}_c = \sqrt{\Delta_c^2 - \lambda^2} = \Delta_c \sqrt{1 - \eta^2}$ and $r = \frac{1}{4}\ln(\frac{1 + \eta}{1 - \eta})$.
Here, we set $\lambda = \eta \Delta_c$ and assume $\eta < 1$.
In the following, we choose $\phi_L = \pi$.
The quadratic coupling term $H_I = 2g_0 A^{\dag}A Q^2$ can be transformed into
\begin{eqnarray}
\widetilde{H}_I = [2 g_s \widetilde{A}^{\dag}\widetilde{A} + g_p(\widetilde{A}^{\dag 2} + \widetilde{A}^2) + 2g_m]Q^2,
\end{eqnarray}
with $g_s = g_0\cosh(2r)$, $g_p = g_0\sinh(2r)$, and $g_m = g_0\sinh^2(r)$.
The last term in the above equation, $2g_m Q^2$, changes the frequency of the mechanical mode.
If the optomechanical coupling is linear with $H_I' = g_0 A^{\dag}A Q$, then the interaction Hamiltonian
is transformed into $\widetilde{H}_I = [2 g_s \widetilde{A^{\dag}}\widetilde{A} + g_p(\widetilde{A}^{\dag 2} + \widetilde{A}^2) + 2g_m]Q$.
Thus, a force is introduced by the parametric amplification. In fact, in Ref. \cite{Lv2015}, a constant force $F$ is applied on
the mechanical mode to cancel the term $2g_m Q$.
However, for the quadratic coupling case in the present work, the constant force $F$ is not necessary.

The Hamiltonian after the squeezing transformation is transformed into
\begin{eqnarray}
\widetilde{H}_0 &=& \widetilde{\Delta}_c \widetilde{A}^{\dag}\widetilde{A}
+ \frac{\widetilde{\omega}_m}{2}(Q^2 + P^2)
+ 2\widetilde{g}_s \widetilde{A}^{\dag}\widetilde{A}Q^2 \nonumber\\
&& + \widetilde{g}_p (\widetilde{A}^{\dag 2} + \widetilde{A}^2) Q^2,
\end{eqnarray}
where $\widetilde{\omega}_m = \omega_m \sqrt{1 + 4g_0\sinh^2(r)/\omega_m}$, $\widetilde{g}_s = (\omega_m/\widetilde{\omega}_m)g_s$,
and $\widetilde{g}_p = (\omega_m/\widetilde{\omega}_m)g_p$.
It is worth noting that the parametric interaction of the above equation can be adjusted be parameters
$\Delta_c$ and $\eta$. In particular, if $\widetilde{\Delta}_c \gg g_p, \widetilde{\omega}_m$, then we
can safely neglect all terms that oscillate with very high frequencies $2\widetilde{\Delta}_c \pm 2 \widetilde{\omega}_m$.
Using the rotating wave approximation, we obtain a standard quadratic optomechanical Hamiltonian
\begin{eqnarray}
\widetilde{H}_0 &=& \widetilde{\Delta}_c \widetilde{A}^{\dag}\widetilde{A} + \frac{\widetilde{\omega}_m}{2}(Q^2 + P^2)
+ 2\widetilde{g}_s \widetilde{A}^{\dag}\widetilde{A}Q^2. \label{vH0}
\end{eqnarray}

Several observations can be made about the above Hamiltonian. First, the effective detuning $\widetilde{\Delta}_c = \Delta_c\sqrt{1 - \eta^2}$ is determined by the detuning $\Delta_c$ and paramter $\eta$.
The pump field on the nonlinear media changes the frequency of the mechanical mode since
$\widetilde{\omega}_m = \omega_m\sqrt{1 + 4 g_0\sinh(r)^2/\omega_m}$.
In optomechanical systems, the single-photon coupling strength is much smaller than the frequency of mechanical mode, i.e., $g_0 \ll \omega_m$.
In addition, in the present work, we choose $\eta < 0.9999$. Thus, $4g_0\sinh^2(r)/\omega_m \ll 1$,
$\widetilde{\omega}_m \approx \omega_m$, $\widetilde{g}_p \approx g_p$, and $\widetilde{g}_s \approx g_s$.
Second, the effective coupling constant
$g_s$ could be much larger than the single-photon quadratic coupling constant $g_0$, that is,
$g_s = g_0\cosh(2r) = g_0/\sqrt{1 - \eta^2} \gg g_0$ if we choose $\eta \rightarrow 1$ (we assume $\eta < 1$).
Thus, the quadratic coupling strength can be significantly enhanced as we expected.
In addition, if we apply a strong driving field on the cavity, the quadratic coupling strength can be
amplified further as we will see later. Finally, detuned parametric drives have been employed in optomechanical systems
in experiments \cite{Szorkovszky2013,Andrews2015}. Therefore, the present proposal can be
implemented with current available optomechanical technology.

\subsection{Strong optical driving field}
In order to enhance quadratic coupling constant further, we apply a strong driving field on the cavity with the Hamiltonian
\begin{eqnarray}
H_{opt-dri} = \Omega(e^{-i\omega_L t} A^{\dag} + e^{i\omega_L t} A),
\end{eqnarray}
where $\Omega$ and $\omega_L$ are the amplitude and frequency of the strong optical driving field, respectively.
In the rotating reference frame with $U_1 = \exp(-i\omega_L t A^{\dag}A)$ as in Eq.(5), $H_{opt-dri}$ can be rewritten as
\begin{eqnarray}
H_{opt-dri} = \Omega(A^{\dag} + A).
\end{eqnarray}
After the squeezing transformation $A = \cosh(r) \widetilde{A} - e^{-i\phi_L} \sinh(r) \widetilde{A}^{\dag}$
with $\phi_L = \pi$, the above Hamiltonian is transformed into
\begin{eqnarray}
\widetilde{H}_{opt-dri} = \widetilde{\Omega} (\widetilde{A}^{\dag} + \widetilde{A}), \label{vHoptdri}
\end{eqnarray}
with $\widetilde{\Omega} = \Omega e^{r}$. This implies the amplitude of the optical driving field can be
amplified exponentially.

\subsection{Effective Hamiltonian}
Suppose the mechanical mode is driven by a field with Hamiltonian
\begin{eqnarray}
H_{mec-dri} = 2\sqrt{2} E \cos{(\omega_d t)} Q, \label{Hmecdri}
\end{eqnarray}
where $E$ and $\omega_d$ are the amplitude and frequency of the mechanical driving field, respectively.

Combining Eqs.(\ref{vH0}), (\ref{vHoptdri}), and (\ref{Hmecdri}), we obtain the total Hamiltonian
\begin{eqnarray}
\widetilde{H}_{tot} &=& \widetilde{\Delta}_c \widetilde{A}^{\dag}\widetilde{A} + \frac{\widetilde{\omega}_m}{2}(Q^2 + P^2)
+ 2\widetilde{g}_s \widetilde{A^{\dag}}\widetilde{A}Q^2 \nonumber\\
 && + \widetilde{\Omega} (\widetilde{A}^{\dag} + \widetilde{A}) + 2\sqrt{2} E \cos{(\omega_d t)} Q.
\end{eqnarray}
Using the standard linearization procedure of cavity optomechanical systems, we can
replace all operators as $O \rightarrow O_s + \delta O$, where $O_s$ is the steady state mean value of operator
$O$ and $\delta O$ is a small quantum fluctuations. More precisely, $\widetilde{A} \rightarrow \widetilde{A}_s + a$,
$Q \rightarrow Q_s + q$, and $P \rightarrow P_s + p$. Similar to Refs. \cite{Xu2018,Xie2017}, we have $Q_s = P_s = 0$ and $\widetilde{A}_s = -2i\widetilde{\Omega}/(\kappa + 2i\widetilde{\Delta}_c)$. The Hamiltonian can be expressed as
\begin{eqnarray}
\widetilde{H}'_{tot} &=& \widetilde{\Delta}_c a^{\dag}a + \widetilde{\omega}_m b^{\dag}b + \widetilde{g}_s (|\widetilde{A}_s|^2 + a^{\dag}a)(b^{\dag} + b)^2\nonumber\\
&& + \widetilde{g}_s(\widetilde{A}_s a^{\dag} + \widetilde{A}_s^*a)(b^{\dag} + b)^2\nonumber\\
&& + 2E\cos(\omega_d t)(b^{\dag} + b).
\end{eqnarray}
Here, we have used $q \equiv (b^{\dag} + b)/\sqrt{2}$ and $p \equiv i(b^{\dag} - b)/\sqrt{2}$.
We assume the optical driving field is strong enough so that $|\widetilde{A}_s|^2 \gg \langle a^{\dag}a \rangle$
and $\widetilde{g}_s a^{\dag}a(b^{\dag} + b)^2$ can be neglected safely. Without loss of generality, we assume $\widetilde{A}_s$
to be real throughout this paper.

In the rotating frame defined by $U_2 = \exp{[-i\omega_dt(2a^{\dag}a + b^{\dag}b)]}$, under the rotating wave approximation
by safely neglected all terms oscillating with high frequencies $\pm 2\omega_d$ and $\pm 4\omega_d$, the effective
Hamiltonian is
\begin{eqnarray}
\widetilde{H}_{eff} &=& \Delta_a a^{\dag}a + \Delta_b b^{\dag}b + g_{eff}(a^{\dag}b^2 + a b^{\dag 2}) \nonumber\\
&& + E(b^{\dag} + b), \label{Heff}
\end{eqnarray}
where $\Delta_a = \widetilde{\Delta}_c - 2\omega_d = \Delta_c\sqrt{1 - \eta^2} - 2\omega_d$,
$\Delta_b = \omega_{m,eff} - \omega_d = (\widetilde{\omega}_m + 2\widetilde{g}_s \widetilde{A}_s^2) - \omega_d$.
The effective coupling strength is $g_{eff} = \widetilde{g}_s \widetilde{A}_s \approx g_s \widetilde{A}_s  = \mathcal{A}_{tot} g_0$.
Here, the total amplification of coupling strength $\mathcal{A}_{tot}$ is
\begin{eqnarray}
\mathcal{A}_{tot} &=& \mathcal{A}_{OPA} \mathcal{A}_{opt-dri},\\
\mathcal{A}_{OPA} &=& \frac{1}{\sqrt{1 - \eta^2}},\\
\mathcal{A}_{opt-dri} &=& \widetilde{A}_s = -2i\widetilde{\Omega}/(\kappa + 2i\widetilde{\Delta}_c).
\end{eqnarray}

\begin{figure}[tbp]
\centering {\scalebox{0.6}[0.6]{\includegraphics{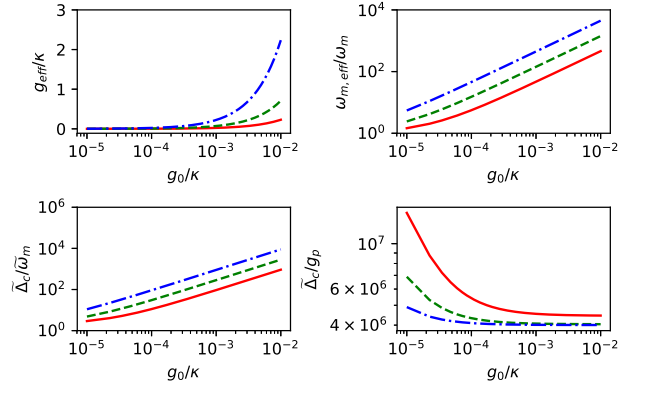}}}
\caption{The quantities $g_{eff}/\kappa$, $\omega_{m,eff}/\kappa$, $\widetilde{\Delta}_c/\widetilde{\omega}_m$, and
$\widetilde{\Delta}_c/g_p$ are plotted as functions of $g_0/\kappa$ for $\eta = 0.9$ (red solid lines), $\eta = 0.99$ (green dashed lines),
and $\eta = 0.999$ (blue dash dot lines). The parameters are $\omega_m = 100\kappa$ and $\widetilde{A}_s = 1000$.
Throughout this work, we consider the resonant case with $\Delta_a = 2\Delta_b$. If $\omega_m$, $g_0$, $\eta$, and $\widetilde{A}_s$ are given, then $\Delta_c$ is determined by Eq.(\ref{Deltac}). Using the relation $\widetilde{\Delta}_c = \Delta_c \sqrt{1 - \eta^2}$, we can calculate
$\widetilde{\Delta}_c$. Note that we assume $\eta < 1$.
} \label{fig1}
\end{figure}

Some remarks must be made now. First, the total amplification of the coupling strength between optical mode
and mechanical mode $\mathcal{A}_{tot}$ is determined by two amplifications. One amplification
is introduced by the OPA. The other is introduced by strong optical driving field. Most importantly,
$\mathcal{A}_{tot} = \mathcal{A}_{OPA} \mathcal{A}_{opt-dri}$ and the coupling constant can be remarkably increased
which is very useful in experiments. For instance, if $\widetilde{A}_s = 1000$ and $\eta = 0.99$, then
the total amplification is more than 7000.
We would like to point the effective coupling constant $g_{eff}$
could be larger than the decay rate of cavity $\kappa$ if we choose appropriate parameters $\eta$ and $\widetilde{A}_s$.
Second, in the derivation of the effective Hamiltonian of Eq.(\ref{Heff}),
we have used the rotating wave approximation ($\widetilde{\Delta}_c \gg g_p, \widetilde{\omega}_m$) and assumed the
optical driving field is strong enough ($\widetilde{A}_s^2 \gg \langle a^{\dag}a \rangle$).
These conditions are satisfied in the discussions of the following sections. Third, in the present work, we consider the
resonant case $\Delta_a = 2\Delta_b$. Given $\omega_m$, $g_0$, $\eta$, and $\widetilde{A}_s$, the resonant condition can be satisfied
by controlling $\Delta_c$ with
\begin{eqnarray}
\Delta_c = [2\sqrt{1 - \eta^2} \widetilde{\omega}_m + 4g_0\widetilde{A}^2_s]/(1 - \eta^2). \label{Deltac}
\end{eqnarray}
In experiments, detuned parametric drives have been employed in optomechanical systems
\cite{Szorkovszky2013,Andrews2015}. Therefore, the present proposal can be
implemented with current available optomechanical technology.
Finally, the mechanical driving field is used to control the second-order correlations, squeezing, and
stationary entanglement of the model. If the mechanical driving filed is weak, there are photon blockade
and phonon blockade even when the quadratic single-photon optomechanical coupling is much smaller than the decay rate of cavity.
The sub-Poissionian window of time-delayed second-order correlations can be controlled by the weak mechanical driving field.
If we increase the amplitude of the driving field, then there are squeezing and stationary entanglement which
can also be tuned by the parameter $E$.

In Fig. 1, we plot the parameters $g_{eff}/\kappa$, $\omega_{m,eff}/\kappa$, $\widetilde{\Delta}_c/\widetilde{\omega}_m$, and
$\widetilde{\Delta}_c/g_p$ as functions of $g_0/\kappa$ for different values of $\eta$.
From the upper left panel of this figure, we see the effective coupling between optical and mechanical modes can be significantly
enhanced by the OPA and strong optical driving field. In fact, $g_{eff}$ could be larger than the decay rate of cavity $\kappa$
if we choose $\eta$ and $\widetilde{A}_s$ appropriately as we can see from the blue dash-dot line of the upper left panel.
When the cavity is driven by a strong driving field, the effective frequency of the mechanical mode
$\omega_{m, eff} = \widetilde{\omega}_m + 2g_s \widetilde{A}^2_s$
can be much larger than the original mechanical frequency $\omega_m$ as one can see from the
upper right panel of Fig. 1. The lower panels of this figure show the conditions
$\widetilde{\Delta}_c \gg \widetilde{\omega}_m, g_p$ are satisfied and the rotating wave
approximation in the derivation of Eq. (\ref{vH0}) is valid.

\section{Second-order correlations}
In this section, we investigate the second-order correlations using the effective Hamiltonian of Eq. (\ref{Heff}).
The second-order correlations in the steady state ($t \rightarrow \infty$) are defined as \cite{Bowen2015}
\begin{eqnarray}
g^{(2)}_{aa}(\tau) &=& \langle a^{\dag}(t)a^{\dag}(t + \tau)a(t + \tau)a(t)\rangle/n^2_a(t),\\
g^{(2)}_{bb}(\tau) &=& \langle b^{\dag}(t)b^{\dag}(t + \tau)b(t + \tau)b(t)\rangle/n^2_b(t),\\
g^{(2)}_{ab}(\tau) &=& \langle a^{\dag}(t)b^{\dag}(t + \tau)b(t + \tau)a(t)\rangle/[n_a(t)n_b(t)],
\end{eqnarray}
where $n_{a}(t) = \langle a^{\dag}(t)a(t) \rangle$ and $n_{b}(t) = \langle b^{\dag}(t)b(t) \rangle$.
Here, without loss of generality, we assume $\tau \geq 0 $.
The equal-time second-order correlation $g^{(2)}_{ij}(0) > 1$ ($g^{(2)}_{ij}(0) < 1$) corresponds to bunching (antibunching).
The perfect photon or phonon blockade effect is observed when $g^{(2)}_{ij}(0) \rightarrow 0$.
Physically, the absorbtion of a photon changes the energy spectrum of an optomechanical system and
the mechanical oscillator is detuned from the cavity. As a result, the probability of absorbing a second photon is
suppressed.

\subsection{Equal-time correlations and photon and phonon blockades}
\subsubsection{Analytical results}
We first discuss the equal-time correlations in a truncated Fock state basis.
If the mechanical driving field is not very strong, then the mean photon and phonon numbers are small.
Thus, we can expand the wave function of the whole system in the few-photon and few-phonon subspace as
\begin{eqnarray}
|\Psi(t)\rangle &=& c_{00}|00\rangle + c_{01} |01\rangle + c_{02}|02\rangle + c_{10}|10\rangle + c_{03}|03\rangle\nonumber\\
&&  + c_{11}|11\rangle + c_{04}|04\rangle + c_{12}|12\rangle + c_{20}|20\rangle,
\end{eqnarray}
where $|nm\rangle \equiv |n\rangle_{photon} \otimes |m\rangle_{phonon}$. Note that the coefficients should
satisfy the conditions $c_{00} \gg c_{01} \gg c_{02}, c_{10} \gg c_{03}, c_{11} \gg c_{04}, c_{12}, c_{20}$
and $c_{00} \approx 1$

The dissipations of cavity and mechanical modes can be taken into accounted by an effective non-Hermitian Hamiltonian
\begin{eqnarray}
\widetilde{H}_{eff}' = \widetilde{H}_{eff} - i\frac{\kappa}{2} a^{\dag}a - i\frac{\gamma}{2}b^{\dag}b, \label{Heff-non-Hermitian}
\end{eqnarray}
where $\kappa$ and $\gamma$ are decay rates of cavity and mechanical modes, respectively.
Substituting the wave function $|\Psi(t)\rangle$ into the Schr\"{o}dinger equation $\frac{d|\Psi(t)\rangle}{dt} = -i \widetilde{H}_{eff}'|\Psi(t)\rangle$ with the non-Hermitian Hamiltonian in Eq.(\ref{Heff-non-Hermitian}) and $\Delta_a = \Delta_b = 0$, we obtain the equation of motion for coefficients $c_{ij}$. If we take $\frac{d c_{ij}}{dt} = 0$, then the steady state values of
coefficients can be obtained. We do not write out them explicitly here since they are too long.

In the case of $c_{00} \gg c_{01} \gg c_{02}, c_{10} \gg c_{03}, c_{11} \gg c_{04}, c_{12}, c_{20}$ and $c_{00} \approx 1$,
the equal-time second-order correlations are
\begin{widetext}
\begin{eqnarray}
g^{(2)}_{aa}(0) &\approx& \frac{2|c_{20}|^2}{|c_{10}|^4}
\approx \frac{\gamma^2(4g_{eff}^2 + \kappa \gamma)^2(-8g^2_{eff} + \kappa^2 + 3\kappa\gamma + 2\gamma^2)^2}
{[8g_{eff}^2 + \gamma(\kappa + \gamma)]^2[\kappa\gamma(\kappa+2\gamma) + 4g^2_{eff}(3\kappa + 2\gamma)]^2},\label{g2aa}\\
g^{(2)}_{bb}(0) &\approx& \frac{2|c_{02}|^2}{|c_{01}|^4}
\approx \frac{\kappa^2\gamma^2}{(4g^2_{eff} + \kappa\gamma)^2},\label{g2bb}\\
g^{(2)}_{ab}(0) &\approx& \frac{|c_{11}|^2}{|c_{01}c_{10}|^2}
\approx \frac{\gamma^2(\kappa + \gamma)^2}{[8g^2_{eff} + \gamma(\kappa + \gamma)]^2}.\label{g2ab}
\end{eqnarray}
\end{widetext}

In Fig. 2, we plot the analytical results (dashed lines) from Eqs. (\ref{g2aa}-\ref{g2ab}) and
numerical results (solid lines) by solving the master equation of Eq.(\ref{master_eq}).
From these equations, we find $g^{(2)}_{bb}(0)$ and $g^{(2)}_{ab}(t)$ decrease with the increase of
$g_0$. However, for $g^{(2)}_{aa}(0)$, the situation is different.
There is a minimal value of $g^{(2)}_{aa}(0)$ and the corresponding effective coupling constant
is denoted by $g_{eff,opt}$. From Eq.(\ref{g2aa}), we see the optimal effective coupling $g_{eff,opt}$ is
\begin{eqnarray}
g_{eff,opt} \equiv \frac{\widetilde{A}_s}{\sqrt{1 - \eta^2}} g_0 = \sqrt{\frac{\kappa^2 + 3\kappa\gamma + 2\gamma^2}{8}}.
\end{eqnarray}
This is consistent with the numerical results as one can see from Fig. 2.
In the following, we solve the master equation of Eq. (\ref{master_eq}) numerically.

\begin{figure}[tbp]
\centering {\scalebox{0.6}[0.6]{\includegraphics{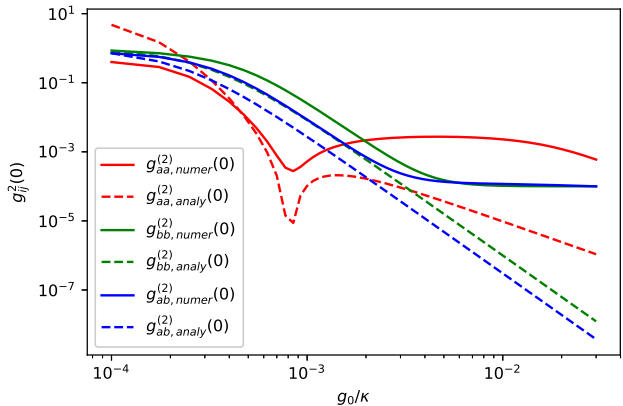}}}
\caption{Equal-time second-order correlations as functions of $g_0/\kappa$ are plotted for analytical results calculated using
Eqs. (\ref{g2aa}-\ref{g2ab}) (dashed lines) and numerical results by solving Eq. (\ref{master_eq}) (solid lines).
The parameters are $\omega_m=100\kappa, \gamma=0.1\kappa, \Delta_a=\Delta_b=0, E=0.05\kappa,
\eta=0, \widetilde{A}_s = 500, N_{th} = 10^{-4}$.
} \label{fig2}
\end{figure}

\subsubsection{Numerical results}

\begin{figure}[tbp]
\centering {\scalebox{0.6}[0.7]{\includegraphics{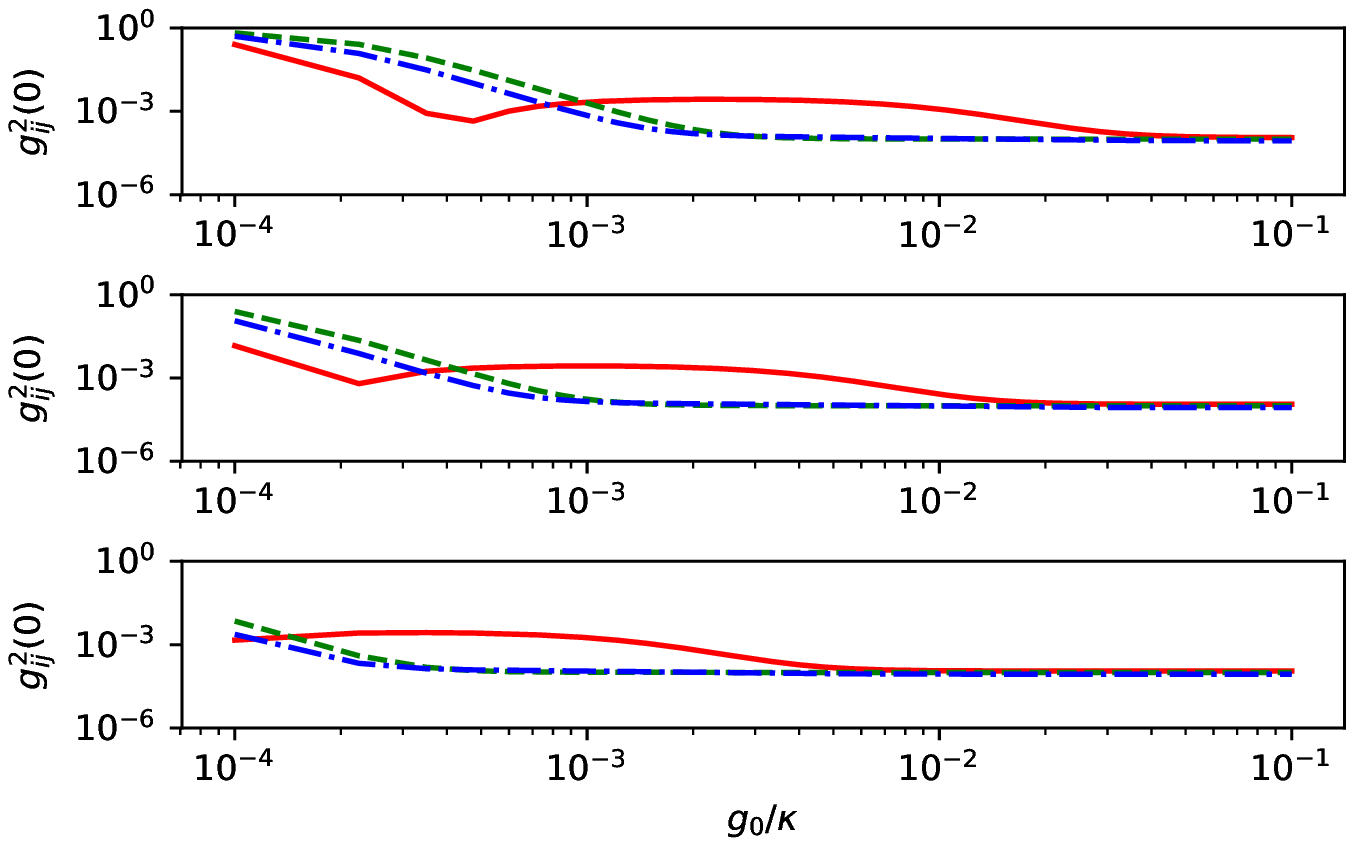}}}
\caption{Equal-time second-order correlations $g^{(2)}_{ij}(0)$ versus $g_0/\kappa$ for $\eta = 0$(upper panel),
$\eta = 0.9$(middle panel), and $\eta = 0.99$(lower panel). $g^{(2)}_{aa}(0)$, $g^{(2)}_{bb}(0)$,
and $g^{(2)}_{ab}(0)$ are represented by red solid, green dashed, and blue dash-dot lines, respectively.
The parameters are $\omega_m=100\kappa, \gamma=0.1\kappa, \Delta_a=\Delta_b=0, E=0.05\kappa,
 \widetilde{A}_s=1000, N_{th}=10^{-4}$.
} \label{fig3}
\end{figure}

\begin{figure}[tbp]
\centering {\scalebox{0.6}[0.7]{\includegraphics{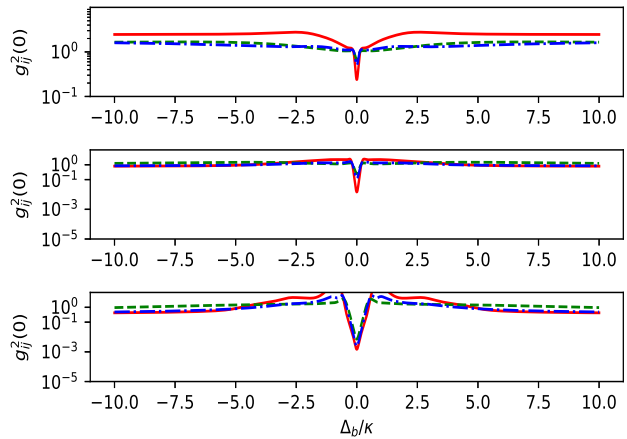}}}
\caption{Equal-time second-order correlations $g^{(2)}_{ij}(0)$ versus $\Delta_b/\kappa$ for $\eta = 0$(upper panel),
$\eta = 0.9$(middle panel), and $\eta = 0.99$(lower panel). $g^{(2)}_{aa}(0)$, $g^{(2)}_{bb}(0)$,
and $g^{(2)}_{ab}(0)$ are represented by red solid, green dashed, and blue dash-dot lines, respectively.
The parameters are $\omega_m=100\kappa, g_0 = 10^{-4}\kappa, \gamma=0.1\kappa, \Delta_a = 2 \Delta_b, E=0.05\kappa,
 \widetilde{A}_s=1000, N_{th}=10^{-4}$.
} \label{fig4}
\end{figure}

As we have pointed out previously, the quadratic coupling strength and the thermal noise of
squeezed cavity mode can be increased simultaneously. Therefore, a squeezed vacuum field must be
introduced to suppress the thermal noise of the squeezed cavity mode completely \cite{Lv2015}.
The present system can also be investigated by numerically solving the following master equation \cite{Lv2015,Yin20182,Xu2018}
\begin{eqnarray}
\frac{d\rho}{dt} &=& -i[\widetilde{H}_{eff}, \rho] + \kappa L[a]\rho \nonumber\\
&&+ \gamma(N_{th} + 1)L[b]\rho + \gamma N_{th} L[b^{\dag}]\rho, \label{master_eq}
\end{eqnarray}
where $N_{th}$ is the mean thermal phonon number and
$L[O]\rho = O\rho O^{\dag} - \frac{1}{2}(O^{\dag} O \rho + \rho O^{\dag}O)$.

In the following, the density matrices of steady state are obtained by solving the master equation numerically.
Then, we calculate equal-time second-order correlations using the density matrices.
The numerical results are plotted as functions of $g_0/\kappa$ in Fig.2 (see solid lines of this figure).
As one can clearly see from Fig.2, there are photon blockade and phonon blockade even when the single-photon quadratic
coupling $g_0$ is much smaller than the decay rate of the cavity $\kappa$. In addition, we find there is photon-phonon blockade, i.e.,
$g^{(2)}_{ab}(0) \ll 1$ if $g_0 > 10^{-3} \kappa$ (see blue lines). This implies that the photons and phonons
are strongly anticorrelated even for weak quadratic coupling $g_0$.
This effect can be used to realize the photon-controlled-phonon (or phonon-controlled-photon) manipulation with
current quantum technologies \cite{Yin20182}.

In order to see the influence of $\eta$ on the second-order correlations more clearly,
we plot equal-time correlations as functions of $g_0/\kappa$  for different values of $\eta$ in Fig.3.
Fixed other parameters, the effective coupling strength $\frac{\widetilde{A}_s}{\sqrt{1 - \eta^2}} g_0$ increases with
$\eta$ (we assume $\eta < 1$ in this paper). Consequently, the second-order correlations decreases with the increase of $\eta$.
For instance, if $g_0 = 10^{-4}\kappa$ and $\eta = 0$, then $g^{(2)}_{ij}(0)$ ($ij = aa, bb, ab$) are about 1 and there are no
photon and phonon blockades. However, in the lower panel of this figure, we see $g^{(2)}_{ij}(0)$ ($ij = aa, bb, ab$)
can be less than $10^{-3}$ if $\eta = 0.99$ even when the original single-photon coupling $g_0$ is much smaller than $\kappa$
($g_0 = 10^{-4}\kappa$).

In Fig. 4, we plot the equal-time correlations as functions of $\Delta_b$ for $\eta = 0$(upper panel),
$\eta = 0.9$ (middle panel), and $\eta = 0.99$ (lower panel). If the nonlinear media are not
pumped ($\eta = 0$), the equal-time correlations are larger than 1 for most detuning $\Delta_b$,
that is, the photons and phonons are bunching. However, if we increase $\eta$, the equal-time correlations
can be remarkably decreased, and there are photon and phonon blockades simultaneously.
The detuning range of phonton and phonon antibunching can be extended by adjusting $\eta$.
The photons and phonons are also strongly anticorrelated as one can see from the blue dash-dot lines in
the middle and lower panels.

\begin{figure}[tbp]
\centering {\scalebox{0.6}[0.7]{\includegraphics{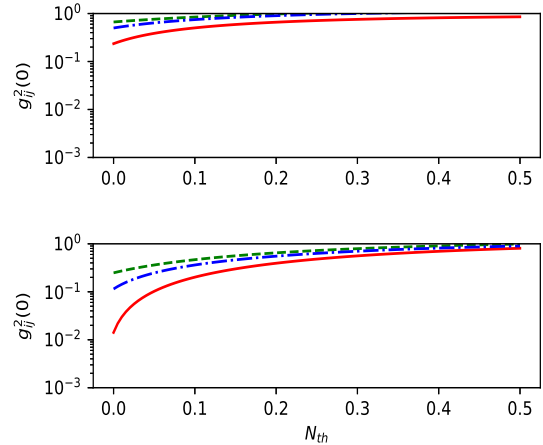}}}
\caption{Equal-time second-order correlations $g^{(2)}_{ij}(0)$ versus $N_{th}$ for $\eta = 0$ (upper panel)
and $\eta = 0.9 $ (lower panel). $g^{(2)}_{aa}(0)$, $g^{(2)}_{bb}(0)$,
and $g^{(2)}_{ab}(0)$ are represented by red solid, green dashed, and blue dash-dot lines, respectively.
The parameters are $\omega_m=100\kappa, g_0 = 10^{-4}\kappa, \gamma=0.1\kappa,
\Delta_a = \Delta_b = 0, E = 0.05\kappa, \widetilde{A}_s=1000$.
} \label{fig5}
\end{figure}

\begin{figure}[tbp]
\centering {\scalebox{0.6}[0.7]{\includegraphics{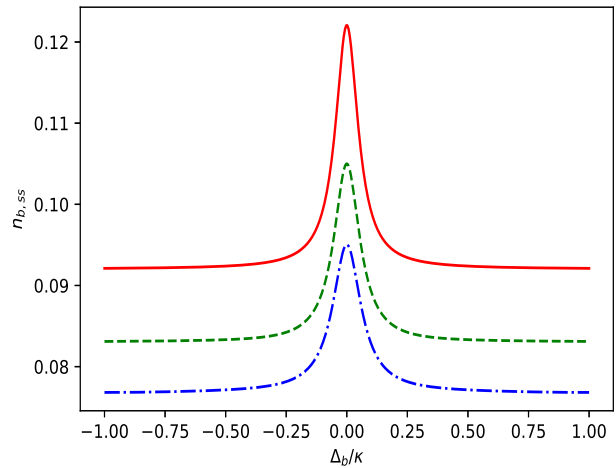}}}
\caption{Mean phonon number $n_{b,ss}$ versus $\Delta_b/\kappa$ for $\eta = 0$ (red solid line),
$\eta = 0.9 $ (green dashed line), and $\eta = 0.99$ (blued dash-dot line).
The parameters are $\omega_m=100\kappa, g_0 = 10^{-4}\kappa, \gamma=0.1\kappa,
\Delta_a = 2\Delta_b, E = 0.01\kappa, \widetilde{A}_s=1000, N_{th} = 0.1$.
} \label{fig6}
\end{figure}

The influence of the mean thermal phonon number $N_{th}$ on the second-order correlations
is displayed in Fig. 5. One can clearly see from Fig. 5 that $g^{(2)}_{ij}(0)$ increases with
$N_{th}$. For example, the correlation of the mechanical mode $g^{(2)}_{bb}(0)$ is larger than 1
for $N_{th} > 0.2$. On the other hand, if the nonlinear media are pumped with amplitude $\eta = 0.9$
then $g^{(2)}_{bb}(0)$ can be decreased remarkably.

In Fig. 6, we display the influence of the amplitude of the driving field applied on the nonlinear media
$\eta$ on the steady-state mean thermal phonon number $n_{b,ss}$. As we expected, $n_{b,ss}$
decreases with the increase of parameter $\eta$.

\subsection{Time-delayed correlations and sub-Poissionian window}
As we have mentioned previously, the UPB scheme was proposed to implement the photon blockade effect
even when the nonlinear energy shifts and driving field are both small \cite{Liew2010}.
But, there is one obstacle in the original UPB scheme.
In fact, in the original UPB, in order to enter the single-photon regime defined by $g^{(2)}(\tau) < 0.5$,
the delayed time $\tau$ must be smaller than the cavity lifetime $1/\kappa$
\cite{Liew2010,Bamba2011,Flayac2013,Flayac2016,Xu2014,Flayac2017}.
Thus, it is difficult to enter the single-photon regime in
experiments since the sub-Poissionian window of the UPB scheme is very small.
In Ref. \cite{Flayac2017}, this obstacle was overcome by employing a mutual driving of the modes and a mixing of the output.

Here, we overcome this obstacle by using a mechanical driving field as we can see from Fig. 7.
The sub-Poissionian windows of time-delayed second-order correlations $g^{(2)}_{ij}(\tau)$ ($ij = aa, bb, ab$)
can be controlled by the amplitude of mechanical driving field $E$. In the case of $E = 0.03\kappa$, the sub-Poissionian window
could be more than $10 /\tau$ (see red solid lines of Fig. 7).

\begin{figure}[tbp]
\centering {\scalebox{0.6}[0.7]{\includegraphics{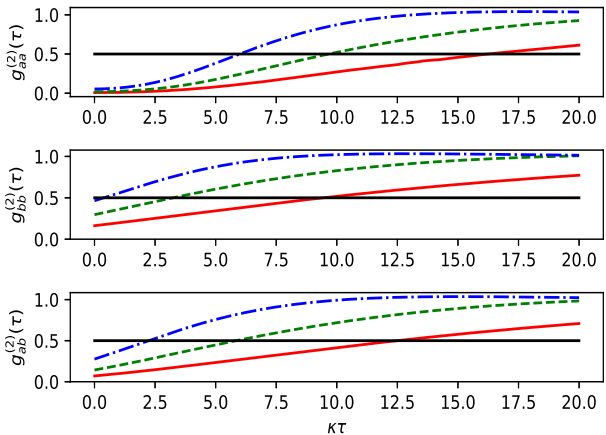}}}
\caption{Time-delayed second-order correlations $g^{(2)}_{ij}(\tau)$ versus $\kappa\tau$ for $E = 0.03\kappa$(red solid lines),
$E = 0.06\kappa$(green dashed lines), and $E = 0.1\kappa$(blue dash-dot lines).
The parameters are $\omega_m=100\kappa, g_0 = 10^{-4}\kappa, \gamma=0.1\kappa, \Delta_a = 2 \Delta_b,
\eta=0.9, \widetilde{A}_s=1000, N_{th}=10^{-4}$.
} \label{fig7}
\end{figure}

\section{Squeezing}

We now turn to study the squeezing of the optical and mechanical modes.
Optomechanical systems are useful in highly sensitive measurement of minuscule displacement.
They are used to probe quantum behavior of macroscopic objects when mechanical oscillators
can be cooled down to their quantum ground state \cite{Bowen2015}.
In order to cool mechanical oscillators down to their quantum ground state, the frequency of
mechanical oscillators must be larger than decay rates of cavities \cite{Bowen2015}.

Squeezing of a mechanical mode plays an essential role in high precision measurements of
position and force of an oscillator since the noise of one quadrature could be less
than that of a coherent state.
The quadratures $X_a$ and $Y_a$ of a field with annihilation operator $a$ are defined by \cite{Agarwal2013}
\begin{eqnarray}
X_a = \frac{a + a^{\dag}}{2}, \quad Y_a = \frac{a - a^{\dag}}{2i}.
\end{eqnarray}
The criterion for nonclassical properties of the field is given by
\begin{eqnarray}
(\Delta X_a)^2 &=& \langle X_a^2\rangle - \langle X_a \rangle^2 < \frac{1}{4},\\
(\Delta Y_a)^2 &=& \langle Y_a^2\rangle - \langle Y_a \rangle^2 < \frac{1}{4}.
\end{eqnarray}
The intermodal quadrature operators are defined by
\begin{eqnarray}
X_{ab} &=& (a + a^{\dag} + b + b^{\dag})/(2\sqrt{2}),\\
Y_{ab} &=& (a - a^{\dag} + b - b^{\dag})/(2i\sqrt{2}),
\end{eqnarray}
and the intermodal squeezing criterion is
\begin{eqnarray}
(\Delta X_{ab})^2 &=& \langle X_{ab}^2\rangle - \langle X_{ab} \rangle^2 < \frac{1}{4},\label{DXab}\\
(\Delta Y_{ab})^2 &=& \langle Y_{ab}^2\rangle - \langle Y_{ab} \rangle^2 < \frac{1}{4}. \label{DYab}
\end{eqnarray}

In Fig. 8, we plot the squeezing of the present model for different values of $\eta$.
From the upper panel, we see there are single mode squeezing and intermodal squeezing only for a small range of detuning $\Delta_b$
if there are no nonlinear media. If the nonlinear media are put into the cavity and pumped by a laser, then the detunig range of squeezing
can be significantly increased as one can clearly see from the middle and lower panels in Fig.8.
For high precision measurements of position, it is crucial to decrease the quantum fluctuations of
position. The squeezing parameters $(\Delta X_b)^2$ and $(\Delta Y_b)^2$ are proportional to the quantum
fluctuations of position and momentum. From the middle and lower panels of Fig.8, we find $(\Delta X_b)^2 < 0.25$ for a large range of
detuning $\Delta_b$ which implies the existence of single mode squeezing.
In addition, there is intermodal squeezing in the present model.

\begin{figure}[tbp]
\centering {\scalebox{0.6}[0.8]{\includegraphics{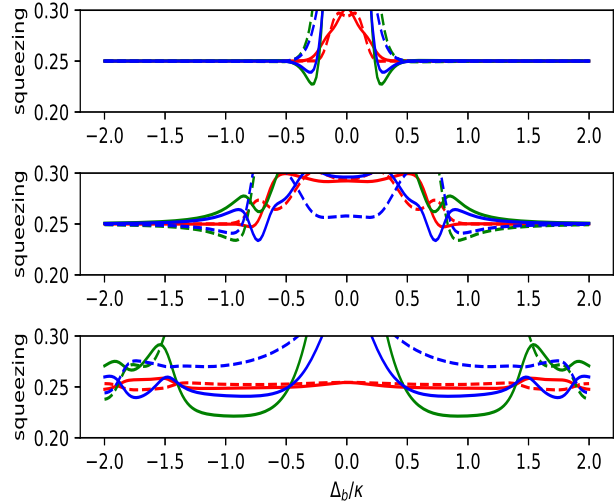}}}
\caption{$(\Delta X_a)^2 \approx (\Delta Y_a)^2 \approx 0.25$ (red solid lines), $(\Delta X_b)^2$ (green solid lines),
$(\Delta Y_b)^2$ (green dashed lines), $(\Delta X_{ab})^2$ (blue lines), and
$(\Delta Y_{ab})^2$ (blue dashed lines) versus detuning $\Delta_b$ for $\eta = 0$ (upper panel), $\eta = 0.99$ (middle panel),
and $\eta = 0.999$ (lower panel).
The parameters are $\omega_m=100\kappa, g_0 = 10^{-4}\kappa, \gamma=0.1\kappa, \Delta_a = 2 \Delta_b,
E = 0.3\kappa, \widetilde{A}_s=1000, N_{th}=10^{-4}$.
} \label{fig8}
\end{figure}

\section{Stationary state entanglement}
In this section, we investigate the steady-state entanglement of the system using two
criteria proposed by by Duan \emph{et al.} \cite{Duan2000}, and Hillery and Zubairy \cite{Hillery2006}.
For any bipartite continuous variable systems, the criterion suggested by
Duan \emph{et al.} is \cite{Duan2000}
\begin{eqnarray}
\mathcal{D}_{ab} = 4(\Delta X_{ab})^2 + 4(\Delta Y_{ab})^2 -2 < 0,
\end{eqnarray}
where $(\Delta X_{ab})^2$ and $(\Delta Y_{ab})^2$ are defined in Eqs. (\ref{DXab}) and (\ref{DYab}).
Note that this criterion is sufficient only for a non-Gaussian state. If $\mathcal{D}_{ab} < 0$,
then we can conclude that the non-Gaussian state is entangled. However, if $\mathcal{D}_{ab} \geq 0$,
we are not sure whether it is entangled or not.
We would like to point out that squeezing can occur at a given time only in one quadrature.
Therefore, the presence of the squeezing of a bipartite system
in general does not guarantee that the bipartite system is entangled.

The other criterion of entanglement for bipartite continuous variable systems was suggested by
Hillery and Zubairy in Ref. \cite{Hillery2006}
\begin{eqnarray}
\mathcal{E}_{1,ab} = \langle a^{\dag}a b^{\dag}b\rangle - |\langle a b^{\dag}\rangle|^2 < 0, \\
\mathcal{E}_{2,ab} = \langle a^{\dag}a \rangle \langle b^{\dag}b\rangle - |\langle a b \rangle|^2 < 0.
\end{eqnarray}

In Fig.9, the two criteria proposed by Duan et.al, and Hillery and Zubairy are plotted as functions of detuning $\Delta_b$ for
different values of $\eta$. In the absence of the nonlinear media, $\mathcal{D}_{ab}$, $\mathcal{E}_{1,ab}$, and $\mathcal{E}_{2,ab}$ are always nonnegative. The situation is very different if the nonlinear media are put into the cavity and pumped by a laser.
For example, in the middle panel, $\mathcal{E}_{1,ab}$ could be negative for $|\Delta_b|/\kappa < 5$. This implies the
cavity and oscillator are entangled at steady state. The detuning range $\Delta_b$ of
entanglement can be extended by increasing $\eta$ as
one can see from the lower panel of Fig.9. In particular, we find the Duan's criterion $\mathcal{D}_{ab}$
fails to detect entanglement, while the criterion proposed by Hillery and Zubairy
$\mathcal{E}_{1,ab}$ can detect the entanglement of the present model.

\begin{figure}[tbp]
\centering {\scalebox{0.6}[0.8]{\includegraphics{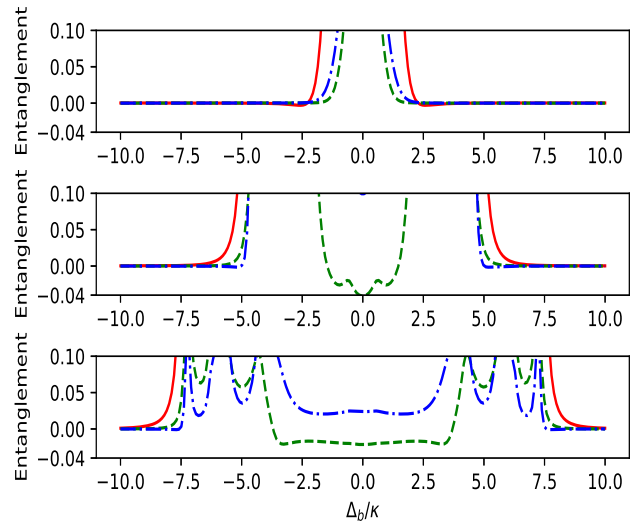}}}
\caption{Entanglement criteria $\mathcal{D}_{ab}$ (red solid lines), $\mathcal{E}_{1,ab}$ (green dashed lines),
and $\mathcal{E}_{2,ab}$ (blue dash-dot lines)
versus detuning $\Delta_b$ for $\eta = 0$ (upper panel), $\eta = 0.997$ (middle panel),
and $\eta = 0.999$ (lower panel).
The parameters are $\omega_m=100\kappa, g_0 = 3\times 10^{-4}\kappa, \gamma=0.1\kappa, \Delta_a = 2 \Delta_b,
E = 2\kappa, \widetilde{A}_s = 1000, N_{th}=10^{-4}$.
} \label{fig9}
\end{figure}

\section{Conclusion}
In the present work, we have proposed a scheme to significantly increase the quadratic coupling
strength of quadratic optomechanics via nonlinear medium and two lasers.
The nonlinear media are pumped by a laser and the optical cavity is pumped by a strong field.
We first derived an effective Hamiltonian of the present system by employing the
rotating wave approximation and squeezing transformation.
Particularly, the total amplification of the quadratic coupling strength is determined by two
amplifications. The first amplification is introduced by the OPA pumped by a laser.
The second amplification is introduced by the strong optical driving field.
We find the quadratic optomechanical coupling constant can be larger than the decay rate of cavity.
This allows us to observe interesting quantum effects such as photon and phonon blockades, single-mode and intermodal
squeezing, and stationary state entanglement of the optical mode and mechanical mode in quadratic optomechanics with
currenttly available optomechanical technology.

Then, we applied a driving field on the mechanical mode and studied the photon blockade and phonon blockade effects.
Our results show that there are photon, phonon, and photon-phonon blockades even when the original quadratic coupling strength is
much smaller than the decay rate of cavity. The sub-Poissionian
window of time-delayed second-order correlations can be adjusted by the weak mechanical driving field.
If we increased the mechanical driving field, there are single mode squeezing and intermodal squeezing.
The range of squeezing could be tuned by the driving field applied on the nonlinear media.
In addition, there is stationary state entanglement between optical and
mechanical modes which can be controlled by the driving field of the nonlinear media.

\section*{Acknowledgement}
This work is supported by
the National Natural Science Foundation of China (Grant Nos. 11047115, 11365009 and 11065007),
the Scientific Research Foundation of Jiangxi (Grant Nos. 20122BAB212008 and 20151BAB202020.)

\end{document}